%% file: 0_main.tex
\documentclass[a4paper]{article}
\pdfoutput=1
\usepackage{INTERSPEECH2020}
\usepackage{array}
\usepackage{hyperref}
\usepackage[utf8]{inputenc}
\newcolumntype{U}[1]{>{\centering\arraybackslash}m{#1}}

\title{Quaternion Neural Networks for Multi-channel Distant Speech Recognition }
\name{Xinchi Qiu$^1$, Titouan Parcollet$^1$, Mirco Ravanelli$^3$, Nicholas Lane$^{1,2}$, Mohamed Morchid$^4$}

\address{
  $^1$University of Oxford, United-Kingdom\\
  $^2$Samsung AI, Cambridge, United-Kingdom\\
  $^3$Mila, Universit\'e de Montréal, Canada\\
  $^4$LIA, Avignon University, France
  }
\email{xinchi.qiu@wolfson.ox.ac.uk}

\begin{document}

\maketitle

\input{1_abstract}

\input{1_introduction}

\input{2_method}
\input{3_exp}

\input{4_conc}

\bibliographystyle{IEEEtran}
\bibliography{mybib}

\end{document}

%% file: 1_abstract.tex
\begin{abstract}
Despite the significant progress in automatic speech recognition (ASR), distant ASR remains challenging due to noise and reverberation. A common approach to mitigate this issue consists of equipping the recording devices with multiple microphones that capture the acoustic scene from different perspectives. These multi-channel audio recordings contain specific internal relations between each signal. In this paper, we propose to capture these inter- and intra- structural dependencies with quaternion neural networks, which can jointly process multiple signals as whole quaternion entities. The quaternion algebra replaces the standard dot product with the Hamilton one, thus offering a simple and elegant way to model dependencies between elements. The quaternion layers are then coupled with a recurrent neural network, which can learn long-term dependencies in the time domain. We show that a quaternion long-short term memory neural network (QLSTM), trained on the concatenated multi-channel speech signals, outperforms equivalent real-valued LSTM on two different tasks of multi-channel distant speech recognition. 
\end{abstract}

%% file: 1_introduction.tex
\section{Introduction}
State-of-the-art speech recognition systems perform reasonably well in close-talking conditions. However, their performance degrades significantly in more realistic distant-talking scenarios, since the signals are corrupted with noise and reverberation \cite{wolfel2009distant, li2015robust, dsp_thesis}. A common approach to improve the robustness of distant speech recognizers relies on the adoption of multiple microphones \cite{brandstein2013microphone,benesty2008microphone}. Multiple microphones, either in the form of arrays or distributed networks, capture different views of an acoustic scene that are combined to improve robustness. 

A common practice is to combine the microphones using signal processing techniques such as beamforming \cite{kellermann}. The goal of beamforming is to achieve spatial selectivity (\textit{i.e.}, privilege the areas where a target speaker is speaking), limiting the effects of both noise and reverberation. One way to perform spatial filtering is provided by the delay-and-sum beamforming, which simply performs a time alignment followed by a sum of the recorded signals \cite{KnappCarter}. More sophisticated techniques are filter-and-sum beamforming \cite{filt_sum}, that filters the signal before summing them up, and super-directive beamforming \cite{super_dir}, which further enhances the target speech by suppressing the contributions of the noise sources from other directions.

An alternative that is gaining significant popularity is End-to-end (E2E) multi-channel ASR \cite{heymann2017beamnet,braun2018multi,tara, unified,7472778,Kim2017}. Here, the core idea is to replace the signal processing part with an end-to-end differentiable neural network, that is jointly trained with the speech recognizer. It will make the speech processing pipeline significantly simpler, and different modules composing the whole system match better with each other.  %This way, the resulting speech processing pipeline is significantly simpler, and the different modules composing the whole system match better with each other. 
The most straightforward approach is concatenating the speech features of the different microphones and feeding them to a neural network \cite{6854663}. However, this approach forces the network to deal with very high-dimensional data, and might thus make learning the complex relationships between microphones difficult due to numerous independent neural parameters. To mitigate this issue, it is common to inject prior knowledge or inductive biases into the model. For instance, \cite{tara} suggested an adaptive neural beamformer that performs filter-and-sum beamforming using learned filters. Similar techniques have been proposed in \cite{unified,7472778}. In all aforementioned works, the microphone combination is not implemented with an arbitrary function, but a restricted pool of functions like beamforming ones. This introduces a regularization effect that helps the convergence of the speech recognizer.
%In \cite{Kim2017}, instead, a neural network that automatically tunes its attention to the most reliable input channel was proposed.

In this paper, we propose a novel approach to model the complex inter- and intra- microphone dependencies that occur in multi-microphone ASR. Our inductive bias relies on the use of quaternion algebra. Quaternions extend complex numbers and define four-dimensional vectors composed of a real part and three imaginary components. The standard dot product is replaced with the Hamilton product that offers a simple and elegant way to learn dependencies across input channels by sharing weights across them. More precisely, Quaternion Neural Networks (QNN) have recently been the object of several research efforts focusing on image processing \cite{parcollet2019survey1,isokawa2003quaternion, parcollet2019quaternion}, 3D sound event detection \cite{comminiello2019quaternion} and single-channel speech recognition \cite{QRNNparcollet2018quaternion}. To the best of our knowledge, our work is the first that proposes the use of quaternions in a multi-microphone speech processing scenario, which is a particularly suitable application. Our approach combines the speech features extracted from different channels into four different dimensions of a set of quaternions (Section \ref{sec:representation}). We then employ a Quaternion Long-Short Term Memory (QLSTM) neural network \cite{QRNNparcollet2018quaternion}. This way, our architecture not only models the latent intra- and inter- microphone correlations with the quaternion algebra, but also jointly learns time-dependencies with recurrent connections.

Our QLSTM achieves promising results on both a simulated version of TIMIT and the DIRHA corpus \cite{7404805}, which are characterized by the presence of significant levels of non-stationary noises and reverberation. In particular, we outperform both a beamforming baseline (15\% relative improvement) and a real-valued model with the same number of parameters (8\% relative improvement). In the interest of reproducibility, we release the code under PyTorch-Kaldi \cite{pytorchkaldi} \footnote{\url{https://github.com/mravanelli/pytorch-kaldi/}}.

%% file: 2_method.tex
\section{Methodology}
\label{sec:met}
This section first describes the quaternion algebra (Section \ref{algebra}) and quaternion long short-term memory neural networks (Section \ref{repre}). Finally, the quaternion representation of multi-channel signals is introduced in Section \ref{sec:representation}. 

\subsection{Quaternion Algebra}\label{algebra}
A quaternion is an extension of a complex number to the four-dimensional space \cite{hamilton1899elements}. A quaternion $Q$ is written as:
\setlength{\belowdisplayskip}{5pt} \setlength{\belowdisplayshortskip}{5pt}
\setlength{\abovedisplayskip}{5pt} \setlength{\abovedisplayshortskip}{5pt}
\begin{equation}
    Q=a+b\textbf{i}+c\textbf{j}+d\textbf{k},
\end{equation}
with $a$, $b$, $c$, and $d$ four real numbers, and $1$, \textbf{i}, \textbf{j}, and \textbf{k} the quaternion unit basis. In a quaternion, $a$ is the real part, while $b\textbf{i}+c\textbf{j}+d\textbf{k}$ with $\textbf{i}^2=\textbf{j}^2=\textbf{k}^2=\textbf{i}\textbf{j}\textbf{k}=-1$ is the imaginary part, or the vector part.  Such definition can be used to describe spatial rotations. In the same manner as complex numbers, the conjugate $Q^*$ of $Q$ is defined as:
\begin{equation}
    Q^* = a - b\textbf{i} -c\textbf{j}-d\textbf{k},
\end{equation}
and a unitary quaternion (\textit{i.e.} whose norm is equal to $1$) is defined as:
\begin{equation}\label{norm}
    Q^\triangleleft = \frac{Q}{\sqrt{a^2+b^2+c^2+d^2}}.
\end{equation}
The Hamilton product between $Q_{1}=a_{1}+b_{1}\textbf{i}+c_{1}\textbf{j}+d_{1}\textbf{k}$ and $Q_{2}=a_{2}+b_{2}\textbf{i}+c_{2}\textbf{j}+d_{2}\textbf{k}$  is determined by the products of the basis elements and the distributive law:
\begin{align}
    Q_{1}\otimes Q_{2}&= (a_{1}a_{2}-b_{1}b_{2}-c_{1}c_{2}-d_{1}d_{2} \nonumber)\\ &+(a_{1}b_{2}+b_{1}a_{2}+c_{1}d_{2}-d_{1}c_{2})\textbf{i}\nonumber\\
    &+(a_{1}c_{2}-b_{1}d_{2}+c_{1}a_{2}+d_{1}b_{2})\textbf{j}\\
    &+(a_{1}d_{2}+b_{1}c_{2}-c_{1}b_{2}+d_{1}a_{2})\textbf{k}. \nonumber
\end{align}
Analogously to complex numbers, quaternions also have a matrix representation defined in a way that quaternion addition and multiplication correspond to a matrix addition and a matrix multiplication. An example of such matrix is:
\begin{equation}
    Q\textsubscript{mat} = 
    \begin{bmatrix}
    a&-b&-c&-d\\
    b&a&-d&c\\
    c&d&a&-b\\
    d&-c&b&a
    \end{bmatrix}
    .
\end{equation}
Following this representation, the Hamilton product can be written as a matrix multiplication as follow:
\begin{equation}\label{eq:mat}
    Q_{1}\otimes Q_{2} = \begin{bmatrix}
    a_{1}&-b_{1}&-c_{1}&-d_{1}\\
    b_{1}&a_{1}&-d_{1}&c_{1}\\
    c_{1}&d_{1}&a_{1}&-b_{1}\\
    d_{1}&-c_{1}&b_{1}&a_{1}
    \end{bmatrix}
    \begin{bmatrix}
    a_{2}\\
    b_{2}\\
    c_{2}\\
    d_{2}\\
    \end{bmatrix} .
\end{equation}
Using the matrix representation of quaternions turns out to be particularly suitable for computations on modern GPUs compared to the less efficient object programming. 

\subsection{Quaternion Long Short-Term Memory Networks}\label{repre}

%QLSTM has been introduced in \cite{QRNNparcollet2018quaternion} to extend the real-valued LSTM to the quaternion domain, to benefit from both the internal relation learning capabilities of quaternion numbers and the natural representation of speech sequences of the LSTM. 
Equivalently to standard LSTM models, a QLSTM consists of a forget gate $f_t$, an input gate $i_t$, a cell input activation vector $\Tilde{C}_t$, a cell state $C_t$ and an output gate $o_t$.
In a QLSTM layer, however, inputs $x$, hidden states $h\textsubscript{t}$, cell states $C\textsubscript{t}$, biases $b$%$b_f$, $b_i$, $b_C$, $b_o$
, and weight parameters $W$ %$W_{fh}$, $W_{fx}$, $W_{ih}$, $W_{ix}$, $W_{Ch}$, $W_{Cx}$, $W_{oh}$, $W_{ox}$ 
are quaternion numbers. All multiplications are thus replaced with the Hamilton product. Different activation functions defined in the quaternion domain can be used \cite{qactivate,parcollet2019survey1}. In this work, we follow the split approach defined as:
\begin{equation}\label{split}
    \alpha(Q) = \alpha(a) + \alpha (b)\textbf{i} +
                \alpha(c)\textbf{j}+\alpha(d)\textbf{k},
\end{equation}
where $\alpha$ is any real-valued activation function (\textit{i.e.} ReLU, Sigmoid, ...). Indeed, fully quaternion-valued activation functions have been demonstrated to be hard to train due to numerous singularities \cite{parcollet2019survey1}. Then, the output layer is commonly defined in the real-valued space to be combined with traditional loss functions (\textit{e.g.} cross-entropy) \cite{qback2} due to the real-valued nature of the labels implied by the considered speech recognition task. Therefore, a QLSTM layer can be summarised with the following equations:
\begin{align} \label{qlstm}
    f_t &= \sigma(W_{fh} \otimes h_{t-1} + W_{fx}\otimes x_t+b_f), \nonumber \\
    i_t &= \sigma(W_{ih} \otimes h_{t-1} + W_{ix}\otimes x_t+b_i), \nonumber \\
    \Tilde{C}_t &= tanh (W_{Ch} \otimes h_{t-1} + W_{Cx}\otimes x_t+b_C), \nonumber \\
    C_t &= f_t \otimes C_{t-1} +i_t \otimes \Tilde{C}_t,  \\
    o_t &= \sigma (W_{oh} \otimes h_{t-1} + W_{ox}\otimes x_t+b_o),\nonumber \\
    h_t &= o_t \otimes tanh(C_t), \nonumber
\end{align}
with two split activations $\sigma$ and $tanh$ as described in Eq. \ref{split}. As shown in \cite{QRNNparcollet2018quaternion}, QLSTM models can be trained following the quaternion-valued backpropagation through time. Finally,  weight initialisation is crucial to train deep neural networks effectively \cite{glorot2010understanding}. Hence, a well-adapted quaternion weight initialisation process \cite{QRNNparcollet2018quaternion, QCNN} is applied. Quaternion neural parameters are sampled with respect to their polar form and a random distribution following common initialization criteria \cite{glorot2010understanding,he2015delving}.

\subsection{Quaternion Representation of Multi-channel Signals}\label{sec:representation}

\begin{figure}[t]
  \centering
  \includegraphics[scale=0.28]{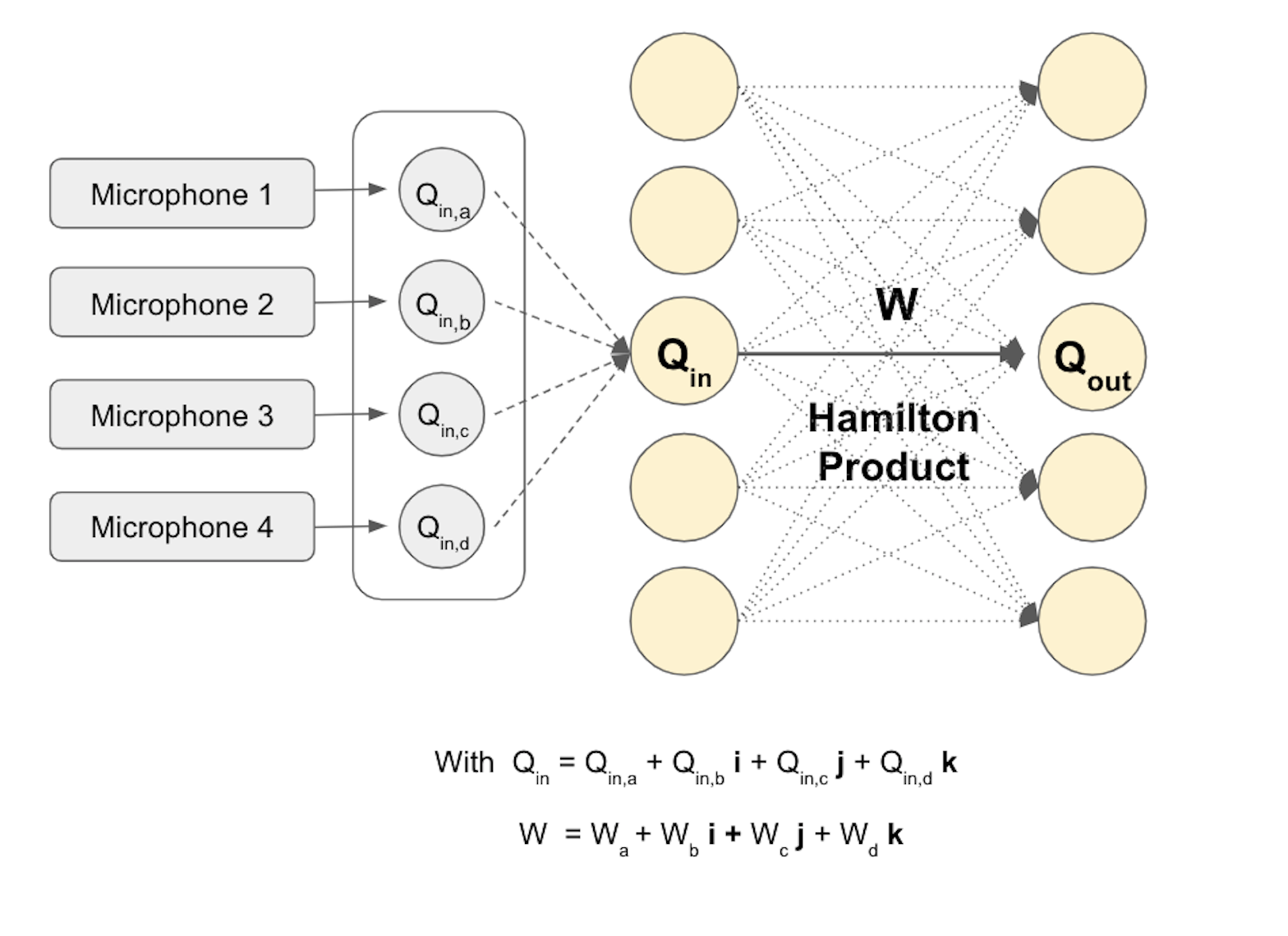}
  \caption{Illustration of the integration of multiple microphones with a quaternion dense layer. Each microphone is encapsulated by one component of a set of quaternions. All the neural parameters are quaternion numbers.}
  \label{fig:graph}
\end{figure}

We propose to use quaternion numbers in a multi-microphone speech processing scenario. More precisely, quaternion numbers offer the possibility to encode up to four microphones (Fig. \ref{fig:graph}). Therefore, common acoustic features (\textit{e.g.} MFCCs, FBANKs, ...) are computed from each microphone signal $M_{1,2,3,4}$, and then concatenated to compose a quaternion as follows:
\begin{equation}
    Q = M_{1,a} +M_{2,b}\textbf{i}+M_{3,b}\textbf{j}+M_{4,b}\textbf{k}
\end{equation}
Internal relations are captured with the specific weight sharing property of the Hamilton product. By using Hamilton products, quaternion-weight components are shared through multiple quaternion-input, creating relations within the elements as demonstrated in \cite{parcollet2019quaternion}. More precisely, real-valued network inputs are treated as a group of uni-dimensional elements that could be related to each other, potentially decorrelating the four microphone signals. Conversely, quaternion networks consider each time frame as an entity of four related elements. Hence, internal relations are naturally captured and learned through the process. Indeed, a small variation in one of the microphone would result in an important change in the internal representation affecting the encoding of the three other microphones.

It is worth noticing that four microphones may be limiting for realistic applications. For instance, the latest CHIME-6 challenge \cite{watanabe2020chime} proposes various recordings obtained from six microphones in different scenarios. This difficulty could be easily avoided by considering these tasks as a special case of higher algebras, such as octonions (eight dimensions) or sedenions (sixteen dimensions). Nevertheless, this paper proposes to first consider four dimensions to evaluate the viability of the application of high-dimensional neural networks for distant and multi-microphone ASR. Finally, quaternion neural networks are known to be more computationally intensive than real-valued neural networks. Indeed, the Hamilton product involves $28$ basic operations compared to $1$ for a standard product. Nonetheless, the training time can be reduced with the matrix representation defined in Eq.\eqref{eq:mat}, and can be drastically improved with simple linear algebra properties \cite{cariow2020fast}. 

%% file: 3_exp.tex
\section{Experimental Protocol}\label{sec:protocol}
A perturbed speech and multi-channel TIMIT \cite{timit} version presented thereafter is first used as a preliminary task to investigate the impact of the Hamilton product. Then, the DIRHA dataset \cite{cristoforetti2014dirha} is used to verify the scalability of the proposed approach to more realistic conditions.  

\subsection{TIMIT Dataset}
The TIMIT corpus contains broadband recordings of $630$ speakers of eight main dialects of American English, each reading ten phonetically rich sentences. The training dataset consists of the standard $3696$ sentences uttered by $462$ speakers, while the testing one consists of $192$ sentences uttered by $24$ speakers. A validation dataset composed of $400$ sentences uttered by $50$ speakers is used for hyper-parameter tuning. 

In our experiments, we created a multi-channel simulated version of TIMIT using the impulse responses measured in \cite{ravanelli2012impulse,ir_selection}\footnote{Perturbation can be re-created following: \url{https://github.com/SHINE-FBK/DIRHA_English_wsj}}. The reference environment is a living room of a real apartment with an average reverberation time $T_{60}$ of $0.7$ seconds. The considered four microphones (\textit{i.e.} $LA2$, $LA3$, $LA4$, $LA5$) are placed on the ceiling of the room. Data are created considering all the different positions, and different positions are used for training and testing data. We also integrate a single-channel signal obtained with delay-and-sum beamforming as a baseline comparison \cite{KnappCarter}. Input features consist of $40$ Mel filters bank energies (FBANK) with no deltas extracted with Kaldi \cite{kaldi}. To show that the obtained gain in performance is independent of the input features, we also propose $13$ MFCC coefficients as an alternative set of features. 

\subsection{DIRHA Dataset}
To validate our model in a more realistic scenario, a set of experiments is also conducted with the larger DIRHA-English corpus \cite{7404805}. Equivalently to the generated TIMIT dataset, the reference context is a domestic environment characterized by the presence of non-stationary noise and acoustic reverberation. Training is based on the original Wall-Street-Journal-5k (WSJ) corpus (\textit{i.e.} consisting of $7138$ sentences uttered by $83$ speakers) contaminated with a set of impulse responses measured in a real apartment \cite{timitrealistic,ravanelli2017contaminated}. Both a real and a simulated dataset are used for testing, each consisting of $409$ WSJ sentences uttered by six native American speakers. Note that a validation set of $310$ WSJ sentences is used for hyper-parameter tuning. Only the first four microphones of the circular array are used in our experiments to fit the quaternion representation. A single-channel signal obtained with delay-and-sum beamforming is also proposed as a baseline comparison \cite{KnappCarter}. It is worth noting that we also used $13$ MFCC coefficients as features in comparison to FBANKs to evaluate the robustness of the model to the input representation.

\subsection{Neural Network Architectures}
We decided to fix the number of neural parameters to $5$M for both LSTM and QLSTM following the models studied in \cite{QRNNparcollet2018quaternion}. Therefore, the QLSTM model is composed of $4$ bidirectional QLSTM layers followed by a linear layer with a softmax activation function for classification. Output labels are the different HMM states of the Kaldi decoder. Each of the QLSTM layers consists of $128$ quaternion nodes. Although there are $128*4=512$ real-valued nodes in total, there are only $128*128*4$ real-valued weight parameters, due to the weight sharing property of quaternion neural networks. The LSTM model is composed of 4 bidirectional LSTM layers of size $290$ (\textit{i.e.} ensuring the same number of neural parameters as the QLSTM) followed by the same linear layer to obtain posterior probabilities. A dropout rate of $0.2$ is applied across all (Q)LSTM layers. Quaternion parameters are initialised with the specific initialisation defined in \cite{QRNNparcollet2018quaternion}, while LSTM parameters are initialised with the Glorot criterion \cite{glorot2010understanding}. 

Training is performed with the RMSPROP optimizer with vanilla hyper-parameters and an initial learning rate of $1.6e^{-3}$ over 24 epochs. The learning-rate is halved every time the loss on the validation set increases, ensuring an optimal convergence. Finally, both LSTM and QLSTM are manually implemented in PyTorch to alleviate any variation due to different implementations.

\section{Results and Discussions}\label{sec:results}

\begin{table*}[ht]
\caption{Results expressed in terms of Phoneme Error Rate (PER) percentage (\textit{i.e} lower is better) of both QLSTM and LSTM models on the TIMIT distant phoneme recognition task with different acoustic features. Results are from an average of $5$ runs.}
\label{tab:timit}
\centering
\begin{tabular}{U{1.75cm}U{4cm}U{1.6cm}U{1.6cm}}
\toprule
\textbf{Models} & \textbf{Signals}  & \textbf{Test (FBANK) }& \textbf{Test (MFCC) }\\
\midrule
QLSTM  & 1 microphone copied & 32.1 $\pm$ 0.02 & 34.2  $\pm$ 0.13 \\
LSTM  & 1 microphone &   32.3 $\pm$ 0.14 & 35.0 $\pm$ 0.23  \\
LSTM  & beamforming    & 31.1 $\pm$ 0.11  & 33.4 $\pm$ 0.07 \\
LSTM    & 4 microphones & 30.2 $\pm$ 0.16  & 32.8 $\pm$ 0.09  \\
\textbf{QLSTM} & \textbf{4 microphones}  & \textbf{28.7 $\pm$ 0.06 } & \textbf{30.4  $\pm$ 0.11 } \\
\bottomrule
\end{tabular}
\end{table*}

\begin{table*}[ht]
  \caption{Results expressed in terms of Word Error Rate (WER) (\textit{i.e} lower is better) of both QLSTM and LSTM based models on the DIRHA dataset with different acoustic features. 'Test Sim.' corresponds to the simulated test set of the corpus, while ``Test Real"  is the set composed of real recordings.  }
  \label{tab:dirha}
  \centering
  \begin{tabular}{U{1.75cm}U{2cm}U{1.6cm}U{1.6cm}U{1.6cm}U{1.6cm}}
    \toprule
    \textbf{ Models}  & \textbf{Signals}& \textbf{Test Real (MFCC)} & \textbf{Test Sim. (MFCC)} & \textbf{Test Real (FBANK)} & \textbf{Test Sim. (FBANK)}\\
    \midrule
    LSTM  & beamforming    & 35.1    & 33.7  & 35.0    & 33.0                     \\
    LSTM  & 4 microphones    & 32.7   & 26.4  & 31.6     &  26.3             \\
    \textbf{QLSTM}     & \textbf{4 microphones}      & \textbf{29.8} &  \textbf{23.8}  & \textbf{29.7}      &  \textbf{23.4}                               \\
    \bottomrule
  \end{tabular}
\end{table*}

The results on the distant multi-channel TIMIT dataset are reported in Table \ref{tab:timit}.
From this comparison, it emerges that QLSTM with four microphones outperforms the other approaches. Our best QLSTM model, in fact, obtains a PER of $28.7\%$ against a PER of $30.2\%$ achieved with a standard real-value LSTM. In both cases, the best performance is obtained with FBANK features. 
%It is worth underlining that our best PER of $28.7\%$ is observed with a QLSTM fed by FBANK features (we achieved a PER of $30.2\%$ for a LSTM with the same input configuration. The same behavior is observed while processing MFCC features, with $30.4\%$ and $32.8\%$ for the QLSTM and LSTM respectively. 
Interestingly, Table \ref{tab:timit} shows that the concatenation of the four input signals with a real-valued LSTM outperforms the  delay-and-sum beamforming approach. %Indeed, an average improvement of $0.75\%$ PER absolute is obtained with the LSTM fed by beamformed signals, increasing to $2.7\%$ absolute with the QLSTM. %Mirco: I don't thins it is good to give to much space of that aspect. 
Similar achievements have already emerged in previous works on multi-channel ASR \cite{6854663} and can be due to the ability of modern neural networks to obtain disentangled and informative representations from noisy inputs. 

We can now investigate in more detail the role played by the quaternion algebra on learning cross-microphone dependencies. One way to do it is to overwrite the quaternion dimensions with the features extracted from the same microphone (see the first row of Table \ref{tab:timit}). In this case, we expect that our QLSTM will fail to learn cross-microphone dependencies, simply because we have a single feature vector replicated multiple times. 
For a fair comparison, the aforementioned experiment is conducted by selecting the best microphone of the array (\textit{i.e.} LA4).

From the first and the second rows of Table \ref{tab:timit}, one can note that both single-channel QLSTM and LSTM perform roughly the same. As expected, in fact, the single-channel QLSTM is not able to model useful dependencies when the quaternion dimensions are dumped with the same feature vector. %There is actually a little advantage for the QLSTM, due to a very slight regularization effect from the quaternion algebra. => Mirco: 0.1 is nothing..
Nonetheless, switching to four-channel signal brings an average PER improvement of $3.6\%$ for the QLSTM compared to $2.1\%$ for the LSTM, showing a higher gain obtained on multiple channels with the QLSTM. This illustrates the ability of QLSTM to better capture latent relations across the different microphones.

To provide some experimental evidence on a more realistic task, we evaluate our model with the DIRHA dataset. The results obtained in Table \ref{tab:dirha} confirm the trend observed with TIMIT. Indeed, Word Error Rates (WER) of $29.8\%$ and $23.8\%$ are obtained for the QLSTM on the real and simulated test sets respectively, compared to $32.7\%$ and $26.4\%$ for the equivalent real-valued LSTM. The same remark holds while feeding our models with FBANK features with a best WER of $29.7$ obtained with the QLSTM compared to $31.6$. As a side note, the accuracies reported on Table \ref{tab:dirha} are slightly worse compared to the ones given in \cite{pytorchkaldi}. Indeed, the latter work includes a specific batch-normalisation that is not applied in our experiments due to the very high complexity of the Quaternion Batch-Normalisation (QBN) introduced in \cite{gaudet2018deep}. As a matter of fact, the current equations of the QBN induce an increase of the VRAM consumption by a factor of $4$. As expected, WER observed on the real test set are also higher than those on the simulated one, due to more complex and realistic perturbations. %An overall gain of $x$ absolute is obtained with the QLSTM compared to the common LSTM. 

As shown in both TIMIT and DIRHA experiments, the performance improvement observed with the QLSTM is independent of the initial acoustic representation, implying that a similar increase of accuracy may be expected with other acoustic features such as fMLLR or PLP. Interestingly, the single-channel beamforming approach gives the worst results among all the investigated methods on both TIMIT and DIRHA.

%We used both FBANKs and MFCCs as described in Section \ref{sec:protocol}. We implement these experiments using the same QLSTM and LSTM networks as before, and all of them take four microphone features as input. From Table \ref{tab:timit}, we can clearly observe that QLSTM perform better than LSTM in both cases, with at least $1.4\%$ improvement. This reinforce the advantage of using quaternion neural network, and also shows that the improvement is independent of input features.

%% file: 4_conc.tex
\section{Conclusion}

\textbf{Summary.} This paper proposed to perform multi-channel speech recognition with an LSTM based on quaternion numbers. Our experiments, conducted on multi-channel TIMIT and DIRHA have shown that: 1) Given the  same number of parameters, our multi-channel QLSTM significantly outperforms an equivalent LSTM network; 2) the performance improvement is observed with different features, implying that a similar increase of accuracy may be expected with others acoustic representations such as fMLLR or PLP; 3) our QLSTM  learns internal latent relations across microphones.
Therefore, the initial intuition that quaternion neural networks are suitable for multi-channel distant automatic speech recognition has been verified.

\noindent
\textbf{Perspectives.} One limitation of the current approach is due to the fact that quaternion neural networks can only deal with four-dimensional input signals. Even though popular devices such as the Microsoft Kinect, or the ReSpeaker are based on 4-microphones arrays, future efforts will focus on generalising this paradigm to an arbitrary number of microphones by considering, for instance, higher dimensional algebras such as octonions and sedenions, or by investigating other methods of weight sharing for multi-channel ASR. Finally, despite recent works on investigating efficient quaternion computations, the current training and inference processes of the QLSTM remain slower than that of a LSTM. Therefore, efforts should be put in developing and implementing faster training procedures.
%Furthermore, specific optimisation methods must be designed to fasten the training procedure of QNNs which remain slower compared to real-valued LSTM.

\section{Acknowledgements}
This work was supported by the EPRSC through MOA (EP/S001530/) and Samsung AI. We would also like to thank Elena Rastorgueva and Renato De Mori for the helpful comments and discussions.

%We would also like to thank many anonymous reviewers for their input throughout the submission process for Interspeech 2020.